\documentclass[notitlepage]{revtex4-1}

\usepackage[utf8]{inputenc}
\usepackage[colorlinks=true,citecolor=blue,linkcolor=blue]{hyperref}
\usepackage[normalem]{ulem}
\usepackage{url}
\usepackage{graphicx,wrapfig,float,slashed,cancel}
\usepackage{amsmath,amssymb,epsfig,graphicx,xcolor,stmaryrd}
\usepackage{bm}
\usepackage{enumitem}
\usepackage{bbm}
\usepackage{multirow}

\definecolor{darkblue}{RGB}{1, 90, 173}

\begin{document}


\title{Investigation of full heavy $ QQQQ'\bar{Q}$ pentaquark candidates}

\author{K.~Azizi}
\thanks{Corresponding author}
\email{ E-mail: kazem.azizi@ut.ac.ir}
\affiliation{Department of Physics, University of Tehran, North Karegar Avenue, Tehran
14395-547, Iran}
\affiliation{Department of Physics, Do\v{g}u\c{s} University, Dudullu-\"{U}mraniye, 34775
Istanbul, T\"{u}rkiye}
\affiliation{School of Particles and Accelerators, Institute for Research in Fundamental Sciences (IPM) P.O. Box 19395-5531, Tehran, Iran}
\author{Y.~Sarac}
\email{E-mail: yasemin.sarac@atilim.edu.tr}
\affiliation{Electrical and Electronics Engineering Department,
Atilim University, 06836 Ankara, T\"{u}rkiye}
\author{H.~Sundu}
\email{ E-mail: hayriyesundu.pamuk@medeniyet.edu.tr}
\affiliation{Department of Physics Engineering, Istanbul Medeniyet University, 34700 Istanbul, T\"{u}rkiye}

\date{\today}

\preprint{}

\begin{abstract}

Recent breakthroughs in research and experimentation have led to the identification of numerous exotic states in particle physics. Each new discovery not only sparks excitement for future findings but also fuels interest in uncovering additional unknown states. Motivated by this perspective and the recent identification of both standard and exotic hadrons  with an increasing number of heavy quarks, this study conducts a spectroscopic analysis of possible  pentaquark candidates  with spin-parity $\frac{1}{2}^-$, and quark content of $cccb\bar{c}$ and $bbbc\bar{b}$. The masses of these states are calculated by considering the relevant Lorentz structures, including $\slashed{p}$ and $\mathbbm{1}$, yielding the following results, respectively: for the $P_{(4cb)}$ state, $m_{P_{(4cb)}} = 11388.30 \pm 107.79$~MeV and $m_{P_{(4cb)}} = 11368.30 \pm 112.68$~MeV, and for the $P_{(4bc)}$ state, $m_{P_{(4bc)}} = 20998.30 \pm 121.52$~MeV and $m_{P_{(4bc)}} = 20990.50 \pm 125.87$~MeV. Additionally, the current coupling constants of these states to the vacuum, which are essential for analyzing their potential decay modes, are also provided in this study.

\end{abstract}


\maketitle

\renewcommand{\thefootnote}{\#\arabic{footnote}}
\setcounter{footnote}{0}
\section{\label{sec:level1}Introduction}\label{intro}

Starting from the first time that the quark model was introduced, hadrons with unusual quark-gluon configurations, that is, those that don't fit the typical definitions of baryons (three quarks or antiquarks) or mesons (a quark-antiquark pair), have attracted considerable attention. Since the existence of such states is allowed by the theory of strong interactions, they were always on the focus of interest. On the other hand, the first indication of their presence came in 2003 with the discovery of the X(3872) state~\cite{Choi:2003ue}. Then, this significant finding was followed by the identification of the many other potential exotic states~\cite{Aaij:2015tga,Aaij:2016ymb,Aaij:2019vzc,LHCb:2020jpq,Zyla:2020zbs,LHCb:2021auc,LHCb:2021vvq,LHCb:2021chn,LHCb:2022ogu}, which are also recorded in the Particle Data Group (PDG)~\cite{ParticleDataGroup:2024cfk}. These discoveries have ignited a wide range of theoretical studies aimed at uncovering their internal structures and properties, as their precise composition remains uncertain. In-depth analyses could potentially lead to the identification of new exotic states. The anticipation of discovering new exotic states calls for thorough investigations and proposals of these potential states, accompanied by detailed insights into their properties, to guide future experimental research. Moreover, these states provide valuable information about the dynamics of the strong interaction, deepening our understanding of it.

Among the examples of the exotic states are the pentaquarks, with their initial discovery reported in 2015 by the LHCb collaboration~\cite{Aaij:2015tga}. The two pentaquark states have arisen from the investigations of decay process $\Lambda_b^0 \rightarrow J/\psi p K^-$. Their masses  and widths were reported to be : $m_{P_c(4380)^+}=4380 \pm8 \pm 29~\mathrm{MeV}$, $\Gamma_{P_c(4380)^+}=205 \pm 18 \pm 86~\mathrm{MeV}$ and $m_{P_c(4450)^+}=4449.8 \pm 1.7 \pm 2.5~\mathrm{MeV}$, $\Gamma_{P_c(4450)^+}= 39 \pm 5 \pm 19~\mathrm{MeV}$~\cite{Aaij:2015tga}. The full amplitude analyses performed in 2016 supported these observations~\cite{Aaij:2016ymb}. With the investigations of updated data in 2019, a new pentaquark state came in sight with mass and width $m_{P_c(4312)^+}=4311.9 \pm 0.7^{ +6.8}_{-0.6}~\mathrm{MeV}$ and $\Gamma_{P_c(4312)^+}=9.8 \pm 2.7 ^{ +3.7}_{-4.5}~\mathrm{MeV}$, respectively, and two thin intersecting crests in the place of the previously  seen  peak of the $P_c(4450)^-$ were reported. Their masses and widths were announced as $m_{P_c(4440)^+}=4440.3 \pm 1.3 ^{+ 4.1}_{-4.7}~\mathrm{MeV}$, $\Gamma_{P_c(4440)^+}= 20.6 \pm 4.9^{+8.7}_{-10.1}~\mathrm{MeV}$ and $m_{P_c(4457)^+}=4457.3 \pm 0.6 ^{+ 4.1}_{-1.7}~\mathrm{MeV}$, $\Gamma_{P_c(4457)^+}= 6.4 \pm 2.0^{+5.7}_{-1.9}~\mathrm{MeV}$~\cite{Aaij:2019vzc}. Later on, the observation of a new pentaquark state with strangeness from the  $J/\psi \Lambda$ invariant mass distribution in $\Xi_b^-\rightarrow J/\psi K^-\Lambda$ decays~\cite{LHCb:2020jpq}, namely the $P_{cs}(4459)^0$, has increased the number of family members, which has mass $m=4458.8 \pm 2.9^{+4.7}_{-1.1}~\mathrm{MeV}$, and width $\Gamma = 17.3 \pm 6.5^{+8.0}_{-5.7}~\mathrm{MeV}$. This observation is followed by the addition of another pentaqauark member to the family from the amplitude studies of $B^{-}\rightarrow J/\psi \Lambda \bar{p}$~\cite{LHCb:2022ogu}, $P_{cs}(4338)$, with the following mass and width values: $4338.2 \pm 0.7\pm 0.4$~MeV and $ 7.0\pm 1.2 \pm 1.3$~MeV, respectively.

The discovery of the pentaquark states has collected theoretical efforts to investigate their various properties. The uncertainties regarding their substructures and quantum numbers make them especially interesting for theoretical research. Studying these states not only enhances our understanding of their nature and composition but also provides information for future experiments aiming at finding similar exotic states. Additionally, their quark configurations, which differ from traditional hadrons composed of three quarks or a quark-antiquark pair, offer important insights into the non-perturbative aspects of quantum chromodynamics (QCD). Due to all these attractive features, various properties of these states were probed deeply via different theoretical approaches and substructure assumptions.  In Refs.~\cite{Chen:2015loa,Chen:2015moa,He:2015cea,Meissner:2015mza,Roca:2015dva,Azizi:2016dhy,Azizi:2018bdv,Azizi:2020ogm,Chen:2020opr,Li:2024wxr,Li:2024jlq,Wang:2023eng,Chen:2020uif,Peng:2020hql,Chen:2020kco,Wang:2019hyc,Wang:2021itn,Wang:2018waa,Liu:2019tjn,Du:2021fmf,Liu:2024ugt}, pentaquarks were assigned to be meson baryon molecular states. They were considered as diquark-diquark-antiquark in Refs.~\cite{Lebed:2015tna,Li:2015gta,Maiani:2015vwa,Anisovich:2015cia,Wang:2015ava,Wang:2015epa,Wang:2015ixb,Ghosh:2015ksa,Wang:2015wsa,Zhang:2017mmw,Wang:2019got,Wang:2020rdh,Ali:2020vee,Wang:2016dzu,Wang:2020eep} and diquark-triquark in Refs.~\cite{Wang:2016dzu,Zhu:2015bba}. Their properties were examined by the topological soliton model~\cite{Scoccola:2015nia} and a version of the D4-D8 model~\cite{Liu:2017xzo} and in Refs~\cite{Guo1,Guo2,Mikhasenko:2015vca,Liu1,Bayar:2016ftu,Co:2024szz} the investigation over their being kinematical effects were presented. Besides the reported pentaquark states, new pentaquark states likely to be observed in the future were also probed in various works~\cite{Liu:2020cmw,Chen:2015sxa,Feijoo:2015kts,Lu:2016roh,Irie:2017qai,Chen:2016ryt,Zhang:2020cdi,Paryev:2020jkp,Gutsche:2019mkg,Azizi:2017bgs,Cao:2019gqo,Azizi:2018dva,Zhang:2020vpz,Wang:2020bjt,Xie:2020ckr,Feijoo:2024qgq,Wang:2024yjp,Liu:2024dlc,Oset:2024fbk,Song:2024yli,Yang:2024okq,Wang:2024brl}.

The advancements in experimental techniques and facilities, along with new discoveries, enhance the expectation of identification of possible new exotic states. Driven by this anticipation, various theoretical models have been applied to explore these states, focusing on their possible structures and quark compositions~\cite{Liu:2020cmw,Chen:2015sxa,Feijoo:2015kts,Lu:2016roh,Irie:2017qai,Chen:2016ryt,Zhang:2020cdi,Paryev:2020jkp,Gutsche:2019mkg,Azizi:2017bgs,Cao:2019gqo,Azizi:2018dva,Zhang:2020vpz,Wang:2020bjt,Xie:2020ckr,Duan:2024uuf,Kong:2024scz,Sharma:2024wpc,Zhang:2023teh,Yan:2023iie,Liu:2023clr,Wang:2023mdj,Wang:2023ael,Yang:2023dzb,Liu:2023oyc,Paryev:2023uhl,Lin:2023iww,Xin:2023gkf,Chen:2023qlx,Zhu:2023hyh,Yan:2023kqf,Sharma:2024pfi,Liu:2024mwn,Roca:2024nsi}. The observation of conventional baryons containing double valence heavy quarks and exotic states composed entirely of heavy quarks~\cite{LHCb:2017iph,LHCb:2018pcs,LHCb:2020bwg,Bouhova-Thacker:2022vnt,Zhang:2022toq,CMS:2023owd,ATLAS:2023bft} have brought up the anticipation of pentaquark resonances with two, three, or even more heavy valence quarks in their inner configurations. This expectation collects the motivations for the investigation of such systems. In this study, we, therefore, focus on pentaquark states composed entirely of heavy quarks, more precisely, four-charm and one-bottom quark, or the reversed case. Similar investigations aiming to identify the properties of full heavy pentaquarks have been conducted in various previous works~\cite{Zhang:2020vpz,Wang:2021xao,Yan:2021glh,An:2020jix,An:2022fvs,Zhang:2023hmg,Yang:2022bfu,Liang:2024met,Alonso-Valero:2024jim,Rashmi:2024ako,Gordillo:2024sem,Azizi:2024ito}. However, further analyses are essential to deepen our understanding of their nature and potential structures, ultimately supporting future experimental efforts. Moreover, such analyses help enhance our understanding of the non-perturbative regime of QCD  and deepen our insight into the dynamics of heavy quarks. To search for these resonances, we employ the QCD sum rule (QCDSR)  method~\cite{Shifman:1978bx,Shifman:1978by,Ioffe81}, which is one of the predictive and powerful non-perturbative approaches and has proven itself in various applications resulting in predictions in nice agreement with experimental findings. The goal of the current analysis is to determine the masses and current coupling constants of fully heavy pentaquark states with spin-parity quantum numbers of $\frac{1}{2}^-$, consisting of four $c(b)$ quarks/antiquark and one $b(c)$ quark. To achieve this, we construct an appropriate interpolating current using the relevant quark fields, taking into account the specified quark content and spin-parity quantum numbers. These states may be detected in future experiments, and thus, the theoretical analysis provided here could assist in guiding such experimental searches by offering crucial insights to inform the investigations.

The paper is organized as follows: Section~\ref{II} outlines the QCDSR utilized to calculate the masses and current coupling constants of the states under study. In Section~\ref{III}, we present the numerical analysis of the results derived from the QCDSR. The final section offers a summary and conclusion of the findings.

\section{Sum rules for physical observables}\label{II}

This section presents the details of the mass and current coupling constant calculations for the considered full heavy pentaquark states with $QQQQ'\bar{Q}$ quark content, where $Q(Q')$ is $c(b)$ or $b(c)$ quarks. From now on we represent these two states as $P_{(4cb)}$ and $P_{(4bc)}$. To obtain the masses and the current coupling constants, the following correlation function is applied:
\begin{equation}
\Pi(p)=i\int d^{4}xe^{ip\cdot
x}\langle 0|\mathcal{T} \{J_{P_{(4QQ')}}(x)\bar{J}_{P_{(4QQ')}}(0)\}|0\rangle,
\label{eq:CorrF1PQ}
\end{equation}
where $\mathcal{T}$ denotes time-ordering operator and $J_{P_{(4QQ')}}$ stands for the all-heavy pentaquark interpolating fields with quark content composed of four $Q$ and one $Q'$ quark, that is the interpolating current of either $P_{(4cb)}$ or $P_{(4bc)}$ states. This interpolating current has the following form:
\begin{eqnarray}
J_{P_{(4QQ')}}&=&[\epsilon^{ijk} Q^{T}_{i}C\gamma_{\mu} Q_{j} \gamma_{5} \gamma^{\mu} Q'_{k}][\bar{Q}_{l} i\gamma_5  Q_{l}].\label{eq:Current}
\end{eqnarray}
In Eq.~(\ref{eq:Current}), $T$ stands for transpose,  $i,~j,~k,~l$ represent the SU(3) color members and  $Q(Q')$, as said,  is either charm  or bottom quark field, and $C$ represents the charge conjugation operator. Taking $Q\equiv c$ and $Q' \equiv b$  or vice-verse, we get the masses and current couplings corresponding to the pentaquark states with quark contents $cccb\bar{c}$ and $bbbc\bar{b}$, namely for $P_{(4cb)}$ and $P_{(4bc)}$ bound-states. 

The QCD sum rule approach involves two main steps:
\begin{itemize}
\item  QCD side: First, the correlator is calculated using the operator product expansion (OPE),  where the calculation is expressed in terms of fundamental QCD parameters, such as the QCD coupling constant, quark masses, and quark-gluon condensates. 
\item Hadronic representation: The same correlator is expressed in terms of hadronic degrees of freedom, including the mass of the hadron and the current coupling constant. This side represents the hadronic or phenomenological side.
\end{itemize}
To match these two sides, a dispersion relation is employed, and the coefficients of the same Lorentz structures are compared. The two sides are then equated, which leads to the extraction of important hadronic quantities like the masses and current coupling constants of the hadrons under consideration in terms of fundamental quantities and vice versa.
In addition to the contributions from the lower states, the calculations also include contributions from higher states and the continuum. To mitigate the effects of these unwanted contributions, Borel transformations are applied to suppress the continuum and higher states. This process helps isolate the physical contributions from the ground state and lowers the impact of higher-energy states, ensuring more accurate results.

For the calculation of the QCD side,  we  use the interpolating current Eq.~(\ref{eq:Current}) explicitly inside the correlation function.  After possible contractions among  the quark fields, making use of Wick's theorem from field theory, the results take the following form:  
\begin{eqnarray}
\Pi^{\mathrm{QCD}}(p)&=&i\int d^4x e^{ip\cdot x}2\epsilon_{abc}\epsilon_{a'b'c'}\gamma^{\mu}\gamma_5 S_{Q'}^{cc'}(x)\gamma_5 \gamma^{\nu} \Big\{-\mathrm{Tr}[\gamma_{\nu}\tilde{S}_Q^{ba'}(x) \gamma_{\mu}S_Q^{ab'}(x)]\mathrm{Tr}[i \gamma_{5} S_Q^{e'e}(-x) i \gamma_{5}S_Q^{ee'}(x)] \nonumber\\
&+&2\mathrm{Tr}[i \gamma_5 S_Q^{ea'}(x)  \gamma_{\nu} \tilde{S}_Q^{ab'}(x)  \gamma_{\mu} S_Q^{be'}(x) i\gamma_5 S_Q^{e'e}(-x) ]\Big\},
\label{eq:QCDSide}
\end{eqnarray}
where $S_Q(x)$ is the heavy quark propagator and $\tilde{S}_Q^{ab}(x)=C S_Q^{abT}(x)C$.  $S_Q(x)$, is given  by
\begin{eqnarray}
S_{Q}^{ab}(x)&=&\frac{i}{(2\pi)^4}\int d^4k e^{-ik \cdot x} \left\{
\frac{\delta_{ab}}{\!\not\!{k}-m_Q}
-\frac{g_sG^{\alpha\beta}_{ab}}{4(k^2-m_Q^2)^2}[\sigma_{\alpha\beta}(\!\not\!{k}+m_Q)+
(\!\not\!{k}+m_Q)\sigma_{\alpha\beta}]\right.\nonumber\\
&&\left.+\frac{\pi^2}{3} \langle \frac{\alpha_sGG}{\pi}\rangle
\delta_{ab}m_Q \frac{k^2+m_Q\!\not\!{k}}{(k^2-m_Q^2)^4}
+\cdots\right\}.
\label{eq:Qpropagator}
\end{eqnarray}
In Eq.~(\ref{eq:Qpropagator}), the $G_{ab}^{\alpha\beta}=G_A^{\alpha\beta}t_{ab}^A$, $GG=G_A^{\alpha\beta}G_A^{\alpha\beta}$, $t^A=\frac{\lambda^A}{2}$ with $\lambda^A $ being the Gell-Mann matrices, $a,~b=1,~2,3$ and  $A=1,~2,\cdots,~8$. 
After inserting the heavy quark propagator into Eq.~(\ref{eq:QCDSide}), we apply the Fourier and  Borel transformations then continuum subtraction supplied by the quark hadron duality assumption. All these operations end up with lengthy final results, therefore we shall not give them here explicitly. The results appear as coefficients of two Lorentz structures, namely $\slashed{p}$ or $\mathbbm{1}$.  We present the final results as:
\begin{eqnarray}
\mathcal{B}\Pi^{\mathrm{QCD}}_i(s_0,M^2)=\int_{(4m_Q+m_{Q'})^2}^{s_0} ds e^{-\frac{s}{M^2}}\rho_i(s),
\label{Eq:Cor:QCD}
\end{eqnarray}
where $\rho_i(s)$  ($ i $ stands for either  $\slashed{p}$ or $\mathbbm{1}$) represent spectral densities attained by taking the imaginary parts of the obtained   results, $\frac{1}{\pi}Im[\Pi_i^{\mathrm{QCD}}]$ either from the coefficients of the Lorentz structure $\slashed{p}$ or $\mathbbm{1}$.

In the hadronic representation calculation, the interpolating currents are used as operators that either create or annihilate the hadrons in question. To proceed, we insert a complete set of hadronic states into the correlation function, ensuring that these states have the same quantum numbers as the interpolating current. This insertion leads to an expression that involves the mass and current coupling constant of the hadron, which we aim to determine. After isolation of the ground state contribution, we get 
\begin{eqnarray}
\Pi^{\mathrm{Had}}(p)= \frac{\langle 0|J|P_{(4QQ')}(p,s)\rangle \langle P_{(4QQ')}(p,s)|\bar{J}|0\rangle}{m_{P_{(4QQ')}}^2-p^2}+\cdots,
\label{eq:hadronicside1}
\end{eqnarray}
where the $\cdots$ represents the contribution of higher Fock states and continuum.  The  one-particle state with momentum $p$ and spin $s$ is represented by $|P_{(4QQ')}(p,s)\rangle$.  The following matrix element is the definition for vacuum to one-particle state through the interpolating  current in terms of the residue or current coupling,  $\lambda_{P_{(4QQ')}}$, and the Dirac spinor, $u(p,s)$:
\begin{eqnarray}
\langle 0|J|P_{(4QQ')}(p,s)\rangle &=& \lambda_{P_{(4QQ')}} u(p,s).
\label{eq:matrixelement}
\end{eqnarray}
Implementing this matrix element in Eq.~(\ref{eq:hadronicside1}) with summation over spin
\begin{eqnarray}\label{Rarita}
\sum_s  u (p,s)  \bar{u} (p,s) &= &\!\not\!{p} + m,
\end{eqnarray}
the hadronic side becomes
\begin{eqnarray}\label{PhyssSide}
\Pi^{\mathrm{Had}}(p)&=&\frac{\lambda^{2}_{P_{(4QQ')}}(\!\not\!{p} + m_{P_{(4QQ')}})}{p^{2}-m_{P_{(4QQ')}}^{2}}+\cdots.
\end{eqnarray}
The Borel transformation with respect to $-p^2$ results in
\begin{eqnarray}
\mathcal{B}\Pi^{\mathrm{Had}}(p)&=&\lambda_{P_{(4QQ')}}^{2} e^{-\frac{m_{P_{(4QQ')}}^{2}}{M^2}}(\!\not\!{p} + m_{P_{(4QQ')}}) +\cdots,
\label{PhyssSideF}
\end{eqnarray}
where $M^2$ is the Borel parameter to be fixed in next chapter.

Considering the coefficient of the same Lorentz structures, the results of the QCD and hadronic sides are matched to achieve the QCD sum rules for the mass and the current coupling constant of the state under quest based on the prescriptions of the method.  One gets the following  two sum rules  for the structures involved:
\begin{eqnarray}
\lambda_{P_{(4QQ')}}^{2} e^{-\frac{m_{P_{(4QQ')}}^{2}}{M^2}}=\mathcal{B}\Pi^{\mathrm{QCD}}_{\slashed{p}}(s_0,M^2),
\label{QCDsumrule}
\end{eqnarray}
and
\begin{eqnarray}
\lambda_{P_{(4QQ')}}^{2} m_{P_{(4QQ')}}e^{-\frac{m_{P_{(4QQ')}}^{2}}{M^2}}=\mathcal{B}\Pi^{\mathrm{QCD}}_\mathbbm{1}(s_0,M^2). 
\label{QCDsumrule1}
\end{eqnarray}
 As we said,  $ \mathcal{B}\Pi^{\mathrm{QCD}}_\mathbbm{i}(s_0,M^2)$ are very  lengthy functions that we don't  present their explicit expressions.

The mass and current coupling are obtained as following after some simple algebra: 
\begin{eqnarray}
m_{P_{(4QQ')}}^2=\frac{\frac{d}{d(-\frac{1}{M^2})}\mathcal{B} \Pi^{\mathrm{QCD}}_{\slashed{p}(\mathbbm{1})}(s_0,M^2)}{\mathcal{B}\Pi^{\mathrm{QCD}}_{\slashed{p}(\mathbbm{1})}(s_0,M^2)}, 
\end{eqnarray}  
and 
\begin{eqnarray}
\lambda_{P_{(4QQ')}}^2=\frac{e^{\frac{m_{P_{(4QQ')}}^2}{M^2}}}{m_{P_{(4QQ')}}}\mathcal{B}\Pi^{\mathrm{QCD}}_{\slashed{p}(\mathbbm{1})}(s_0,M^2).
\end{eqnarray}
This brings us to the next stage in which we use all these results for numerical analyses of the physical quantities under study.

\section{Numeric evaluation of sum rules}\label{III}

\begin{figure} []
\centering
\begin{tabular}{cccc}
\includegraphics[totalheight=5cm,width=7cm]{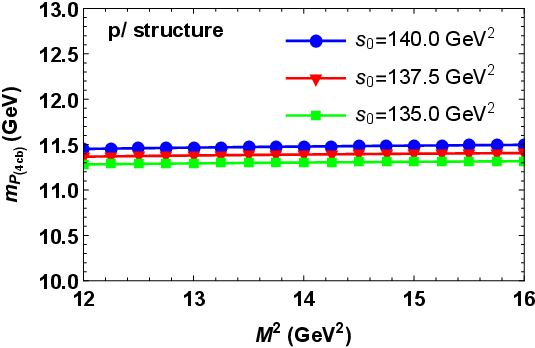} &
\includegraphics[totalheight=5cm,width=7cm]{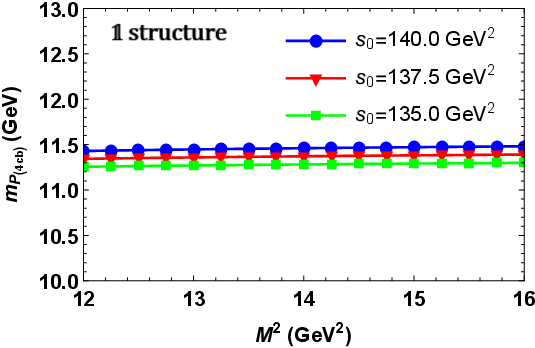} \\
\textbf{(a)}  & \textbf{(b)}  \\[6pt]
\end{tabular}
\begin{tabular}{cccc}
\includegraphics[totalheight=5cm,width=7cm]{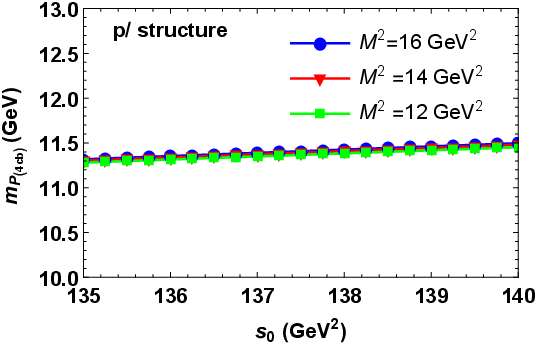} &
\includegraphics[totalheight=5cm,width=7cm]{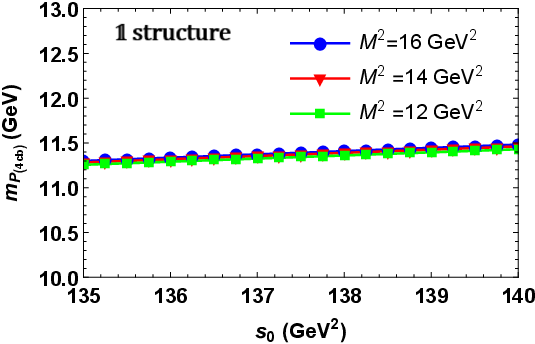} \\
\textbf{(c)}  & \textbf{(d)}  \\[6pt]
\end{tabular}
\caption{ 
(a) The dependence of the mass of the $P_{(4cb)}$ pentaquark state on the Borel parameter $M^2$, calculated for different  values of the  parameter $s_0$, as derived from the Lorentz structure $\slashed{p}$.
(b) The same as (a), but for the  structure $\mathbbm{1}$.
(c) The dependence of the mass of the $P_{(4cb)}$ pentaquark state on the threshold parameter $s_0$, calculated for  different values of the  parameter $M^2$, as derived from the Lorentz structure $\slashed{p}$.
(d) The same as (c) but for the  structure $\mathbbm{1}$.
}
\label{fig:mass4cb1}
\end{figure}

\begin{figure} []
\centering
\begin{tabular}{cccc}
\includegraphics[totalheight=5cm,width=7cm]{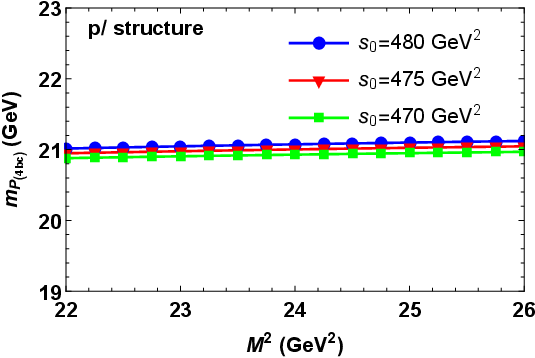} &
\includegraphics[totalheight=5cm,width=7cm]{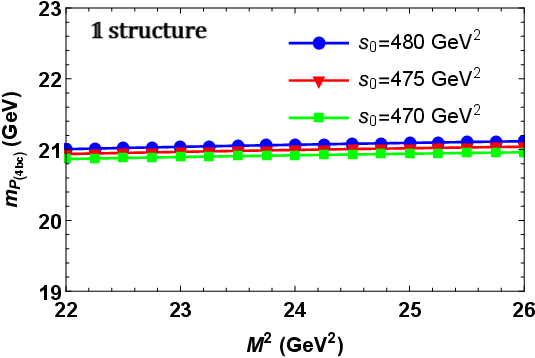} \\
\textbf{(a)}  & \textbf{(b)}  \\[6pt]
\end{tabular}
\begin{tabular}{cccc}
\includegraphics[totalheight=5cm,width=7cm]{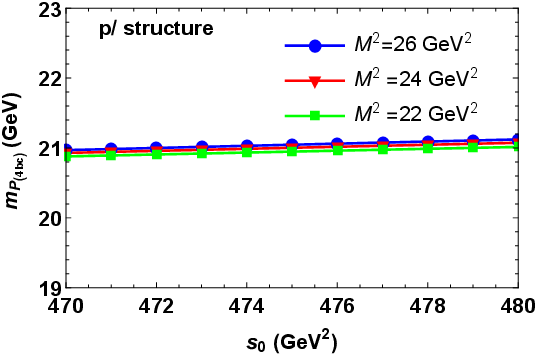} &
\includegraphics[totalheight=5cm,width=7cm]{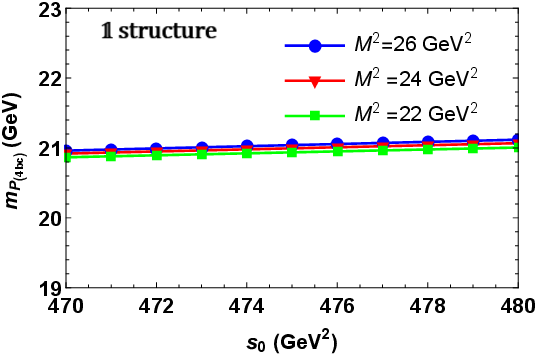} \\
\textbf{(c)}  & \textbf{(d)}  \\[6pt]
\end{tabular}
\caption{ 
(a) The dependence of the mass of the $P_{(4bc)}$ pentaquark state on the Borel parameter $M^2$, calculated for different  values of the parameter $s_0$, as obtained from the Lorentz structure $\slashed{p}$.
(b) The same as (a), but for the  structure $\mathbbm{1}$.
(c) The dependence of the mass of the $P_{(4bc)}$ pentaquark state on the threshold parameter $s_0$, calculated for different values of the  parameter $M^2$, as obtained from the Lorentz structure $\slashed{p}$.
(d) The same as (c), but for the  structure structure $\mathbbm{1}$.
}
\label{fig:mass4bc1}
\end{figure}

\begin{figure} []
\centering
\begin{tabular}{cccc}
\includegraphics[totalheight=5cm,width=7cm]{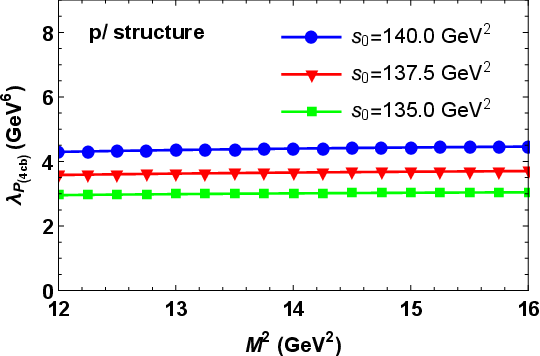} &
\includegraphics[totalheight=5cm,width=7cm]{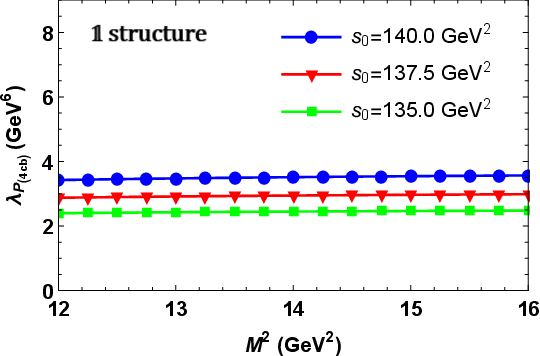} \\
\textbf{(a)}  & \textbf{(b)}  \\[6pt]
\end{tabular}
\begin{tabular}{cccc}
\includegraphics[totalheight=5cm,width=7cm]{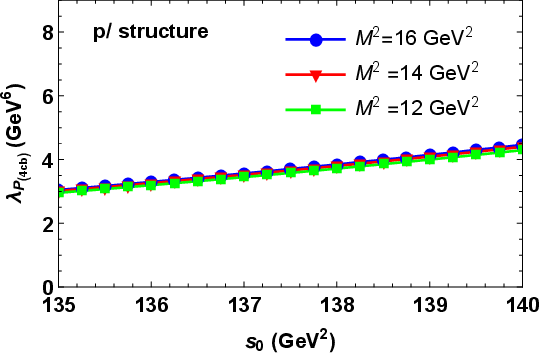} &
\includegraphics[totalheight=5cm,width=7cm]{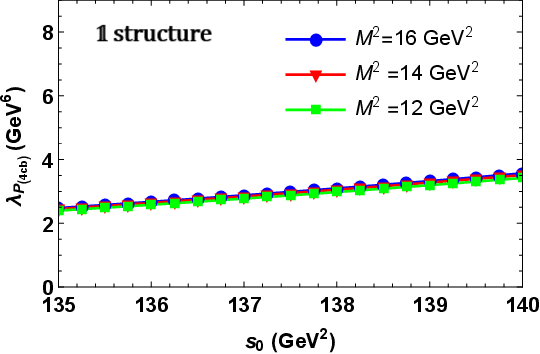} \\
\textbf{(c)}  & \textbf{(d)}  \\[6pt]
\end{tabular}
\caption{ 
(a) The dependence  of the residue or current coupling constant $\lambda_{P_{(4cb)}}$ for the $P_{(4cb)}$ pentaquark as a function of the parameter $M^2$, calculated for different values of  $s_0$, as obtained from the Lorentz structure $\slashed{p}$.
(b) The same as (a), but for the  structure $\mathbbm{1}$.
(c) The dependence  of the residue or current coupling constant $\lambda_{P_{(4cb)}}$ for the $P_{(4cb)}$ pentaquark state as a function of  $s_0$, calculated for different values of  $M^2$, as obtained from the Lorentz structure $\slashed{p}$.
(d) The same as (c), but for the  structure $\mathbbm{1}$.
}
\label{fig:res4cb2}
\end{figure}

\begin{figure} []
\centering
\begin{tabular}{cccc}
\includegraphics[totalheight=5cm,width=7cm]{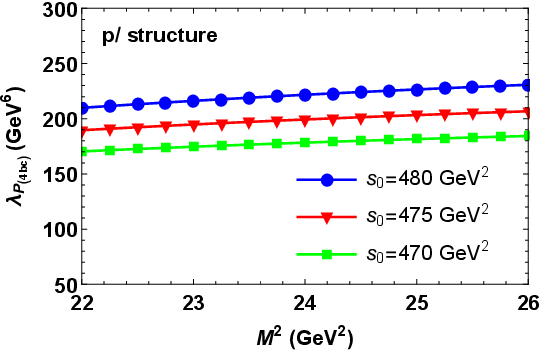} &
\includegraphics[totalheight=5cm,width=7cm]{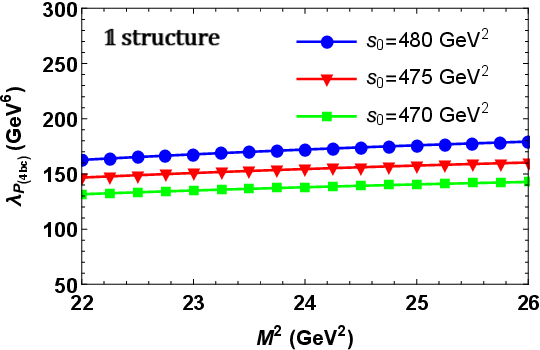} \\
\textbf{(a)}  & \textbf{(b)}  \\[6pt]
\end{tabular}
\begin{tabular}{cccc}
\includegraphics[totalheight=5cm,width=7cm]{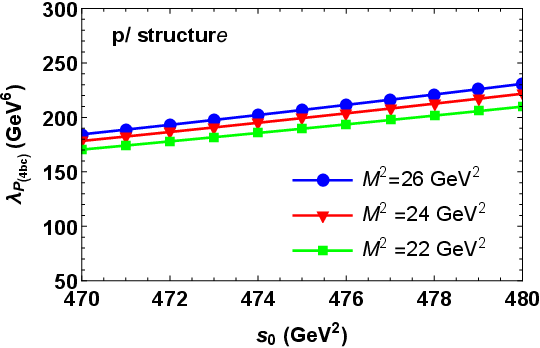} &
\includegraphics[totalheight=5cm,width=7cm]{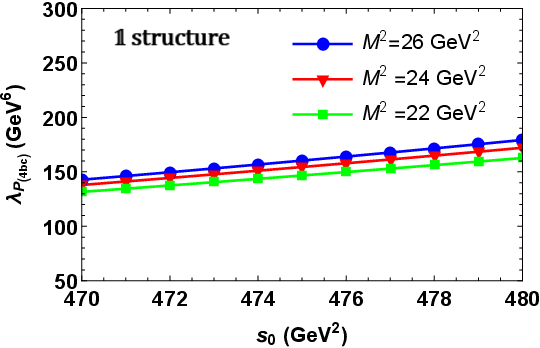} \\
\textbf{(c)}  & \textbf{(d)}  \\[6pt]
\end{tabular}
\caption{ 
(a) The dependence of the current coupling constant $\lambda_{P_{(4bc)}}$ for the $P_{(4bc)}$ pentaquark state on the Borel parameter $M^2$, calculated for different values of the threshold parameter $s_0$, as obtained from the Lorentz structure $\slashed{p}$.
(b) The same as (a), but for the  structure  $\mathbbm{1}$.
(c) The dependence of the current coupling constant $\lambda_{P_{(4bc)}}$ for the $P_{(4bc)}$ pentaquark state on the threshold parameter $s_0$, calculated for different values of the Borel parameter $M^2$, as obtained from the Lorentz structure $\slashed{p}$.
(d) The same as (c), but for the  structure $\mathbbm{1}$.
}
\label{fig:res4bc2}
\end{figure}

The findings from the previous section are utilized with the appropriate input parameters, some of which are outlined in Table~\ref{tab:Inputs}. 
\begin{table}[h!]
\begin{tabular}{|c|c|}
\hline\hline
Parameters & Values \\ \hline\hline
$m_{c}$                                    & $1.2730\pm 0.0046~\mathrm{GeV}$ \cite{ParticleDataGroup:2024cfk}\\
$m_{b}$                                     & $4.18^{+0.04}_{-0.03}~\mathrm{GeV}$ \cite{ParticleDataGroup:2024cfk}\\
$\langle \frac{\alpha_s}{\pi} G^2 \rangle $ & $(0.012\pm0.004)$ $~\mathrm{GeV}^4 $\cite{Belyaev:1982cd}\\
\hline\hline
\end{tabular}%
\caption{The required input parameters for performing the numerical analyses.}
\label{tab:Inputs}
\end{table} 
Nonetheless, these parameters alone are insufficient, as the sum rules involve two additional auxiliary parameters that need to be determined through the analysis of the results. These parameters are the threshold parameter, $s_0$, which emerges from the application of the quark-hadron duality assumption, and the Borel parameter, $M^2$, introduced through the Borel transformation. The values of these parameters are established imposing the  standard guidelines within the QCD sum rule framework. Specifically, the threshold parameter is chosen by considering the energy of the first likely excited state, as it is directly associated with this quantity. As a result, the parameter values are selected based on estimations of the energies corresponding to the potential first excited states of the pentaquarks under study. Accordingly, the analyses are conducted using the following ranges:
\begin{eqnarray}
&135.0~\mbox{GeV}^2 \leq s_0 \leq 140.0~\mbox{GeV}^2 &
\end{eqnarray} 
for $P_{(4cb)}$ state and 
\begin{eqnarray}
&470.0~\mbox{GeV}^2 \leq s_0 \leq 480.0~\mbox{GeV}^2 &
\end{eqnarray} 
for $P_{(4bc)}$ states. To fix the Borel parameters, we consider the convergence of the operator product expansion (OPE), dominance of the pole contribution, and suppression of higher-order and continuum contributions. Taking into account that the higher-dimensional condensates in the expansion do not dominate the sum rule and do not exceed the contribution of the perturbative term we fix the lower limit of the Borel parameters. For the determination of the upper limit, it is required that the ground state's contribution dominates over that of continuum and higher resonance states.  Moreover, it is essential for the results obtained through the QCD sum rule method to exhibit stability within the selected ranges of both the Borel  and the threshold parameters. Considering these criteria, the ranges for the Borel parameters are determined as
\begin{eqnarray}
12.0~\mbox{GeV}^2\leq M^2\leq 16.0~\mbox{GeV}^2,
\end{eqnarray}
for $P_{(4cb)}$ pentaquark state and
\begin{eqnarray}
22.0~\mbox{GeV}^2\leq M^2\leq 26.0~\mbox{GeV}^2,
\end{eqnarray}
for the $P_{(4bc)}$ state. To assess the stability of the results with respect to variations in these parameters, graphs are plotted showing the obtained results as functions of the Borel  and threshold parameters. These graphs, presented in Figures~\ref{fig:mass4cb1}-\ref{fig:res4bc2}, display the masses and current coupling constants calculated for the $cccb\bar{c}$ and $bbbc\bar{b}$ pentaquark states using both Lorentz structures, $\slashed{p}$ and $\mathbbm{1}$. The figures illustrate how these quantities vary with the Borel parameter $M^2$ and the threshold parameter $s_0$ within their respective working intervals. The plots confirm that the results demonstrate the expected stability within the chosen ranges of these parameters.

Subsequently, the determined ranges of the working parameters are employed to extract the numerical values of the masses and current coupling constants, along with their associated uncertainties. These uncertainties stem from the errors in the input parameters as well as the uncertainties introduced by the selection of the working intervals for the auxiliary parameters. The results obtained from both Lorentz structures, $\slashed{p}$ and $\mathbbm{1}$, are summarized in Table~\ref{tab:table2}.
\begin{table}[]
\begin{tabular}{|c|c|c|c|}
\hline
State                        & Structure     & Mass~(MeV) & $\lambda~(\mathrm{GeV}^6)$ \\ \hline \hline
\multirow{2}{*}{$P_{(4cb)}$} & $\slashed{p}$ & $11388.30 \pm 107.79$     & $3.68 \pm 0.73$        \\ \cline{2-4} 
                             & $\mathbbm{1}$ & $11368.30 \pm 112.68$     & $2.96 \pm 0.59$          \\ \hline \hline
\multirow{2}{*}{$P_{(4bc)}$} & $\slashed{p}$ & $20998.30 \pm 121.52$     & $(1.99 \pm 0.30)\times 10^{2}$          \\ \cline{2-4} 
                             & $\mathbbm{1}$ & $20990.50 \pm 125.87$     & $(1.54\pm 0.24) \times 10^{2}$          \\ \hline
\end{tabular}
\caption{The mass and current coupling constant results derived from the QCD sum rules for the $P_{(4cb)}$ and $P_{(4bc)}$ states, analyzed using both Lorentz structures $\slashed{p}$ and $\mathbbm{1}$.}
\label{tab:table2}
\end{table}

\section{Discussion and Concluding remarks}\label{IV}

The number of exotic particles  is continuously growing  as a result of the advancements in experimental methods and analyses techniques.  Every novel  state discovered raises hopes for future findings and highlights the potential for investigating similar states. Inspired by these observed states, theoretical studies have examined systems with quark compositions that differ from those already known, aiming to offer valuable insights or comparison grounds for upcoming experiments. Driven by these advancements, along with recent discoveries of both conventional and exotic states that include a higher number of heavy quarks, this study explores the potential pentaquark states made of charm or bottom quarks with the quark content $cccb\bar{c}$ and $bbbc\bar{b}$ and spin parity $J^P=\frac{1}{2}^-$, which were represented in text as $P_{(4cb)}$ and $P_{(4bc)}$, respectively. The masses for these states were obtained by applying the QCD sum rule method, considering each Lorentz structure, namely $\slashed{p}$ and $\mathbbm{1}$, present in the analyses. The masses were obtained as $m_{P_{(4cb)}}=11388.30 \pm 107.79$~MeV and $m_{P_{(4cb)}}=11368.30 \pm 112.68$~MeV  for $P_{(4cb)}$ state and $m_{P_{(4bc)}}=20998.30 \pm 121.52$~MeV and $m_{P_{(4bc)}}=20990.50 \pm 125.87$~MeV  for $P_{(4bc)}$ state, considering $\slashed{p}$ and $\mathbbm{1}$ Lorentz structures, respectively. We also presented the current coupling constants for these states, which are among the necessary ingredients of the decay width analyses of the considered states.

The pentaquark states with the similar quark content of the present study have been investigated in various works. In Ref.~\citep{Zhang:2023hmg} the masses were obtained for  $P_{cccb\bar{c}}$ state as $m=11.581$~GeV and for $P_{bbbc\bar{b}}$ state as $m=21.493$~GeV. The predictions of the same masses in Ref.~\cite{An:2020jix} are  $m=11175$~MeV and $m=20691$~MeV and in Ref.~\cite{An:2022fvs} $m=11438.2$~MeV and $m=21079.0$~MeV, respectively. Compared to our results the result of Ref.~\cite{Zhang:2023hmg} for $cccb\bar{c}$ and $bbbc\bar{b}$ are larger and that of Ref.~\cite{An:2020jix} are slightly smaller than our corresponding results. On the other hand, our results  are consistent, within the error margins, with the results of Ref.~\cite{An:2022fvs}. 

It is evident that further research is essential to gain a more comprehensive understanding of these states, especially regarding their structure and properties. The results shared here can be helpful in regard to comparison of them with other future studies and also serve as inputs for future inquiries into the interaction processes of these resonances.

\section*{ACKNOWLEDGMENTS}
K. Azizi expresses his gratitude to the Iran National Science Foundation (INSF) for the partial financial support under the Elites Grant No. 4037888. He also extends his thanks to the CERN-TH division for their support and warm hospitality.



\end{document}